\documentclass[aps,
pre,
reprint, 
superscriptaddress
]{revtex4-1}

\usepackage{hyperref}

\usepackage{amssymb}
\usepackage{amsmath}
\usepackage{amsfonts}
\usepackage{graphicx,subfigure}
\usepackage{epstopdf}
\usepackage{hyperref}
\usepackage{color}
\usepackage{eso-pic}
\newcommand{\be}{\begin{equation}}
\newcommand{\ee}{\end{equation}}
\newcommand\numberthis{\addtocounter{equation}{1}\tag{\theequation}}
\newcommand{\PDMP}{P_{\mathrm{DMP}}}

\begin{document}
\title{A message-passing approach for threshold models of behavior in networks}
\date{\today}
\author{Munik Shrestha}
\affiliation{Department of Physics and Astronomy, University of New Mexico, Albuquerque, NM 87131, USA}
\affiliation{Santa Fe Institute, 1399 Hyde Park Road, Santa Fe, NM 87501, USA}
\author{Cristopher Moore}
\affiliation{Santa Fe Institute, 1399 Hyde Park Road, Santa Fe, NM 87501, USA}

\begin{abstract}
We study a simple model of how social behaviors, like trends and opinions, propagate in networks where individuals adopt the trend when they are informed by threshold $T$ neighbors who are adopters. Using a dynamic message-passing algorithm, we develop a tractable and computationally efficient method that provides complete time evolution of each individual's probability of adopting the trend or of the frequency of adopters and non-adopters in any arbitrary networks. We validate the method by comparing it with Monte Carlo based agent simulation in real and synthetic networks and provide an exact analytic scheme for large random networks, where simulation results match well. Our approach is general enough to incorporate non-Markovian processes and to include heterogeneous thresholds and thus can be applied to explore rich sets of complex heterogeneous agent-based models.
\end{abstract}

\maketitle
\section{Introduction}
\label{intro}
Mathematical modeling of epidemics has attracted the interest of researchers from diverse academic disciplines \cite{Bailey75, AM91, Newman1, BMay2011, Strogatz01, MN00a, WS98, Liljeros01, KA01, PV01a, Price65, LM01, DWatts1, JMiller04,CSMF2012, Havlin_Stanley_Interdependent, MJackson}. Epidemics range from outbreaks of infectious disease to the contagion of social behaviors such as trends, memes, fads, political opinions, rumors, innovations, financial decisions, and so on.  In an early study, sociologist Mark Granovetter \cite{Gran78, Gran73} proposed a threshold model, where individuals adopt a behavior when they are informed by at least $T$ of their neighbors. 

We consider a stochastic model similar to Granovetter's with a trend propagating on a network. At each time, an individual has integer valued awareness of a trend ranging from 0 to $T$. Each time an individual is informed by one of its neighbors, this awareness is incremented until it reaches the threshold $T$. At that point, that individual adopts the trend, and starts informing its neighbors about it. We will assume that the network topology is fixed, but our model of information flow (or ``contagion") is probabilistic. Each adopter informs each of its neighbors at a rate $r(\tau)$, where $\tau$ is the time elapsed since it became an adopter. Since $r(\tau)$ may depend on $\tau$, the resulting dynamics can be non-Markovian.  

Given an initial condition, where some individuals have already become adopters, or have done so with some probability, our goal in this paper is to calculate the probability that any given individual $i$ is an adopter (or not an adopter) as a function of time. We can do this by first calculating the probability $P_a^i(t)$ that $i$ has awareness $a$ at time $t$. The probability that $i$ is an adopter is then $P^i_T$. 

Calculating the time evolution of the probability $P_a^i(t)$ is non-trivial as a result of intrinsic nonlinearities in the dynamics. The heterogeneous network interactions between individuals make it even harder. One simple way to estimate these probabilities is to  put on a computational-frequentist hat, simulate the model many times independently by a Monte Carlo agent-based method, and measure in what fraction of these runs each vertex becomes an adopter. Doing this is computationally costly, however, as we are required to perform many independent runs of the simulation

We thus consider the dynamic message passing algorithm (DMP), where we evolve the probabilities $P^i_a(t)$ directly according to certain update equations. Compared to a Monte Carlo simulation that requires many independent runs, we only need to run the DMP algorithm once. In the special case where $T=1$, DMP was recently formulated by Karrer and Newman \cite{KarrNew1} to analytically study non-Markovian dynamics of the Susceptible, Infected, Recovered (SIR) epidemic model of the networks. In an analogy with the SIR model, we sometimes refer to a vertex as \emph{susceptible} if it is not yet an adopter, \emph{infected} if it is an adopter, and \emph{recovered} if it is an adopter but the rate $r(\tau)$ at which it informs its neighbors has dropped to zero.

The underlying idea of dynamic message passing is similar to belief propagation \cite{JPearl1,Decelle2011}, where we use the network structure to update posterior probabilities of the vertices' states. However, unlike belief propagation where we update posterior distributions according to Bayes' rule, the causal structure of information flow is captured directly by the time iteration of DMP. As in belief propagation, the DMP algorithm assumes that the neighbors of each vertex are conditionally independent of each other. As a result, like belief propagation, DMP is exact on trees and approximate on networks with loops, where the conditional independence assumption cannot capture higher order correlations. 

However, as we will see, DMP gives good approximations to the probabilities even on real networks with many loops. We will show this by implementing it in a real social network, specifically Zachary's karate club network~\cite{Zachary}. Although the Zachary's club network contains many loops, the probabilities computed by DMP compare well with those from the Monte Carlo simulation. We present this in Section \ref{karate}. 

In the limit of large random networks in the Erd\H{o}s-R\'enyi model, or networks with a given degree distribution, DMP is asymptotically exact because these networks are locally treelike. In Section  \ref{configuration_network}, we use DMP to obtain the exact results for such random networks in the thermodynamic limit. 

\subsection{Related Work} 

There are many related studies that consider what fraction of vertices eventually become adopters if each neighbor informs them with probability $p$. The set of eventual adopters are the ones who have at least $T$ neighbors who are also adopters. This is reminiscent of the model commonly studied in statistical physics as $k$-core (or bootstrap) percolation. The $k$-core is the maximal induced subgraph in the network, such that each vertex has at least $k$ other neighbors in the subgraph.     

By deleting each edge with probability $1-p$ independently, we can ask whether the resulting diluted network in the thermodynamic limit contains an extensive $k$-core in the ensemble of similarly prepared networks. Interestingly for $k \ge 3$, the emergence of a $k$-core in random networks is a first-order (discontinuous) phase transition in the sense that when it first appears it covers a finite fraction of the network \cite{Pittel91}. An early work on $k$-core percolation was on the Bethe lattice in the context of magnetic systems \cite{Chalupa79}. Recently, it has been used in studies of the Ising model and nucleation \cite{Cerf2011, Cerf2010}, analysis of zero temperature jamming transitions \cite{Jamming2006}, and in a bootstrap percolation model in square lattices and random graphs \cite{Aizenman88, BootGNP12, Amini, DWatts1, BaxDorog}.

 
\section{Message passing approach} \label{message_passing}

We now formulate the dynamic message passing (DMP) technique for the threshold model described in Section \ref{intro}. We define the message $U_{i\leftarrow j}(t)$ as the probability that vertex $j$ has \emph{not}  informed $i$ about the trend by time $t$. If we have $U_{i \leftarrow j}(t)$ for all neighboring pairs $i$, $j$, we will be able to calculate the marginal probability $P^i_{a}(t)$ that $i$ has awareness $a$ at time $t$, i.e. that it has been informed by $a$ of its neighbors. We focus on initial conditions where each vertex is either an adopter or has awareness zero. So given that $i$ is not an initial adopter, 
\be\label{Pai}
P^i_{a}(t) = \sum_{\substack{\Theta \subseteq \partial i\\ |\Theta| =a}}  \prod_{j \in \Theta} (1-U_{i\leftarrow j}(t))   \prod_{j \in \partial i \setminus \Theta } U_{i\leftarrow j}(t).  
\ee
Here, $\partial i$ is the set of $i$'s neighbors, and $\Theta$ ranges over all subsets of $\partial i$ of size $a$. Note the conditional independence assumption in Equation \eqref{Pai}, where we assume that the events that $j$ has informed (or not informed) $i$ are independent. Given that $i$ is not an initial adopter, the probability $P^i_{S}(t)$ that the vertex $i$ is susceptible at time $t$, i.e. its awareness is less than $T$ at time $t$,  is then
\be\label{PS_sum}
P^i_{S}(t) =\sum_a^{T-1} P^i_{a}(t). 
\ee
Equivalently, 
\be\label{PSi}
P^i_{S}(t)=\sum_{ \substack{ \Theta \subseteq \partial i \\ |\Theta| <T}}  \prod_{j \in \Theta} (1-U_{i\leftarrow j}(t))   \prod_{j \in \partial i \setminus \Theta} U_{i\leftarrow j}(t).
\ee
We can see that this expression is easy to generalize to the case where each individual has its own threshold $T_i$. For instance, we could set $T_i$ to some fraction of $i$'s degree. We could also assume a probabilistic threshold $T_i$ for each $i$ drawn from some distribution $P(T_i)$ and take an average over the threshold in Equation \eqref{Pai}. We can also capture the case where $i$ initially has awareness $a_i$ by setting $T_i=T-a_i$. However, for simplicity, we assume that every individual has the same threshold, and everyone starts with an initial awareness of $0$ or $T$. 

Given $P^i_{S}(t)$, we note that $i$ is an adopter if it is at the root of a $T$-ary tree, whose nodes are mapped onto the vertices of the network, such that 1) the leaves of the tree are initial adopters, 2) the $T$ children of each tree node are mapped to distinct vertices, 3) none of the paths from the root to the leaves backtracks; that is, an edge $(u,v)$ cannot be immediately followed by the edge $(v,u)$, and 4) the trend is successfully transmitted along each edge of this tree.

To capture the information flow that the message $U_{i\leftarrow j}(t)$ represents, we define $P_S^{j \setminus i}(t)$, which is the probability that $j$ would be susceptible at time $t$ if $i$ were absent from the network. Alternately, this is the probability that $j$ is susceptible at time $t$ if we ignore the possibility of $j$ being informed of the trend by $i$. In removing the vertex $i$ (or ignoring the flow of information to $j$ from $i$), we bring the information flow to $i$ based on the information or messages that neighbor $j$ receives from $j$'s other neighbors. We thus avoid the ``echo-chamber" effect, where $i$ informs $j$, and $j$ informs $i$ back, and so on. 

In an analogy with the cavity method of statistical physics, we call $P_S^{j \setminus i}(t)$ the cavity probability that $j$ is susceptible given that $i$ is in a noninteracting ``cavity state''. Hence, using Equation \eqref{PSi}, if $j$ was not an initial adopter, then $P_S^{j \setminus i}$ can be written as
\begin{align*} 
 P_S^{j \setminus i}(t) = \sum_{ \substack{\Theta \subseteq \partial j\setminus i \\ | \Theta | <T} }   ~\prod_{\ell \in \Theta} (1-U_{j\leftarrow \ell}(t))    \prod_{  \ell \in \partial j \setminus  \{ \Theta, i \} } & U_{j\leftarrow \ell}(t).  \numberthis \label {cavity_PS} 
 \end{align*}
Note that initially $ P_S^{j \setminus i}(0)= P_S^{j}(0)$, since the initial probability that $j$ is an an adopter does not depend on $i$. Similarly, the cavity rate $p_I^{j\setminus i}(t)$ at which $j$ becomes an adopter at time $t$, if it was not an adopter initially, is then
 \be\label{cavity_rate}
 p_I^{j\setminus i}(t)= -\frac{dP_S^{j \setminus i}(t)}{dt}.
 \ee
 
It is convenient to define $f(\tau)$ as the rate at which $j$ \emph{first} informs $i$ at time $t$, if $j$ became an adopter at time $t'=t-\tau$. In particular, if $j$ informs $i$ at a rate $r(\tau)$, then $f(\tau)= r(\tau) e^{-\int_0^\tau d\tau' r(\tau')}$ is the rate at which $j$ informs $i$ for the first time at time $t$. Note that $f(\tau)$ might not be normalized, since the probability $p = \int_0^\infty d\tau f(\tau)$ that $j$ ever informs $i$ may be less than $1$.  By letting $f(\tau)$ depend arbitrarily on the time $\tau$ since $j$ became an adopter, we can handle both Markovian and non-Markovian models.  In particular, if an adopter inform its neighbors at some constant rate $\beta$, and if it ``recovers" with rate $\gamma$ as in the SIR model, after which it no longer informs its neighbors about the trend, we see that 
\be\label{transmission}
f(\tau)=\beta e^{-(\gamma+\beta) \tau}. 
\ee
Note also that we can let $f(\tau)$ depend on $i$ and $j$, giving arbitrary inhomogeneous rates at which individuals inform each other; we do not pursue this here.
 
Although we have defined the messages and shown how they allow us to calculate the probabilities $P_a^i(t)$, we have not yet shown how to calculate the messages themselves. 

So, let us now calculate the messages $U_{i\leftarrow j}(t)$. The rate at which $U_{i\leftarrow j}(t)$ decreases at time $t$ is the rate at which $j$ informs $i$ for the first time at time $t$. This happens in two ways. If $j$ was an initial adopter, it informs $i$ for the first time at time $t$ at the rate $f(t)$. Or, if $j$ was initially susceptible, $j$ becomes an adopter at some time $t'=t-\tau$, and informs $i$ for the first time at the rate $f(t-t')$ at time $t$. Integrating this over $t'$ up to time $t$, we see that $j$ will inform $i$ for the first time at the rate $\int_0^t dt'f(t-t')p_I^{j \setminus i}(t')$. Combining these two cases with Equation \eqref {cavity_rate}, the rate at which the message $U_{i\leftarrow j}(t)$ decreases at time $t$ is thus given by
 
{\small\begin{align*}
-\frac{dU_{i\leftarrow j}(t)}{dt} &= f(t) [1-P_S^{j}(0)] + P_S^{j}(0) \int_0^t dt'f(t-t')p_I^{j \setminus i}(t') \\& 
 =  f(t) [1-P_S^{j}(0)] -P_S^{j}(0) \int_0^t dt' f(t-t') \frac{ dP_S^{j \setminus i}(t')}{dt'}. 
\numberthis \label {mainUijdiff}
\end{align*}}Integrating by parts gives
\begin{align*}
\frac{dU_{i\leftarrow j}(t)}{dt} = -f(t)&+f(0)P_S^j(0) P_S^{j\setminus i}(t) \\& 
 +  P_S^{j}(0) \int_0^t dt' P_S^{j \setminus i}(t') \frac{df(t-t')}{dt}.
\numberthis \label {mainUijdiff2}
\end{align*}
 One may check that the solution of \eqref{mainUijdiff2} is 
\begin{align*}
U_{i\leftarrow j}(t)  =  1-  \int_0^t d\tau f(\tau)+ P_{S}^{j}(0) \int_0^t d\tau f(\tau)  P_{S}^{j \setminus i}(t-\tau)   \numberthis \label {mainUij}.
\end{align*}
We can explain this expression, as in \cite{KarrNew1}, as follows. The term $1-  \int_0^t d\tau f(\tau)$ is the probability that the elapsed time $\tau$, after which $j$ informs $i$ for the first time, is greater than the absolute time $t$, i.e. $\tau >t$. In this case, $i$ is not informed by $j$, even if $j$ became an adopter before time $t$. The second term is the probability that $i$ would have been informed at time $t$ if $j$ had been an adopter at time $t-\tau$, but that $j$ was not yet an adopter at that time.

Note however that Equation \eqref{mainUijdiff2} is an integro-differential equation, so numerically integrating it can be computationally costly.  It is possible to numerically integrate \eqref{mainUij}, or, for particular functions $f(\tau)$, we can transform  \eqref{mainUijdiff2} into an ordinary differential equation. For example if we plug $f(\tau)$ from \eqref{transmission} and integrate the last term in  \eqref{mainUijdiff2} by parts, we obtain
\begin{align*}
\frac{dU_{i\leftarrow j}(t)}{dt} = -\beta  U_{i\leftarrow j}(t)+\gamma  (1- U_{i\leftarrow j}(t))  +  \beta P_S^{j}(0) P_S^{j \setminus i}(t)
\numberthis \label {mainUijdiff3}
\end{align*}
 So, given the initial conditions $U_{i\leftarrow j}(0)$ and $P^i_S(0)$, we numerically integrate this or \eqref{mainUijdiff2} to  compute $P^i_a(t)$, $P^i_S(t)$, and $P^i_T(t)$ using \eqref{Pai} and \eqref{PSi} respectively. 

\section{Message passing vs Monte Carlo simulation in Real Networks} \label{karate}
\begin{figure*}
\centering
\mbox{\subfigure{\includegraphics[width=3.5in]{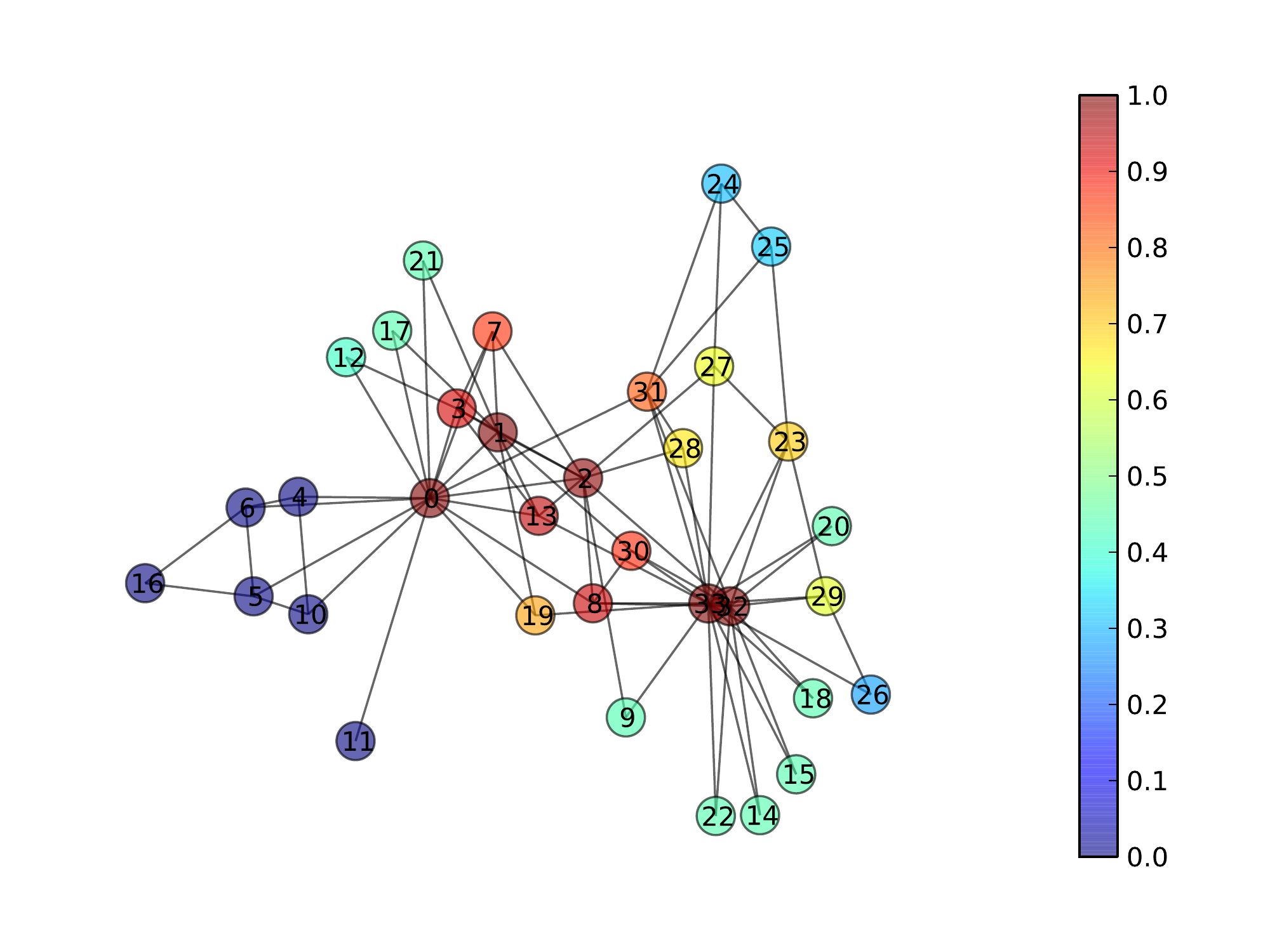}}\quad
\subfigure{\includegraphics[width=3.0in]{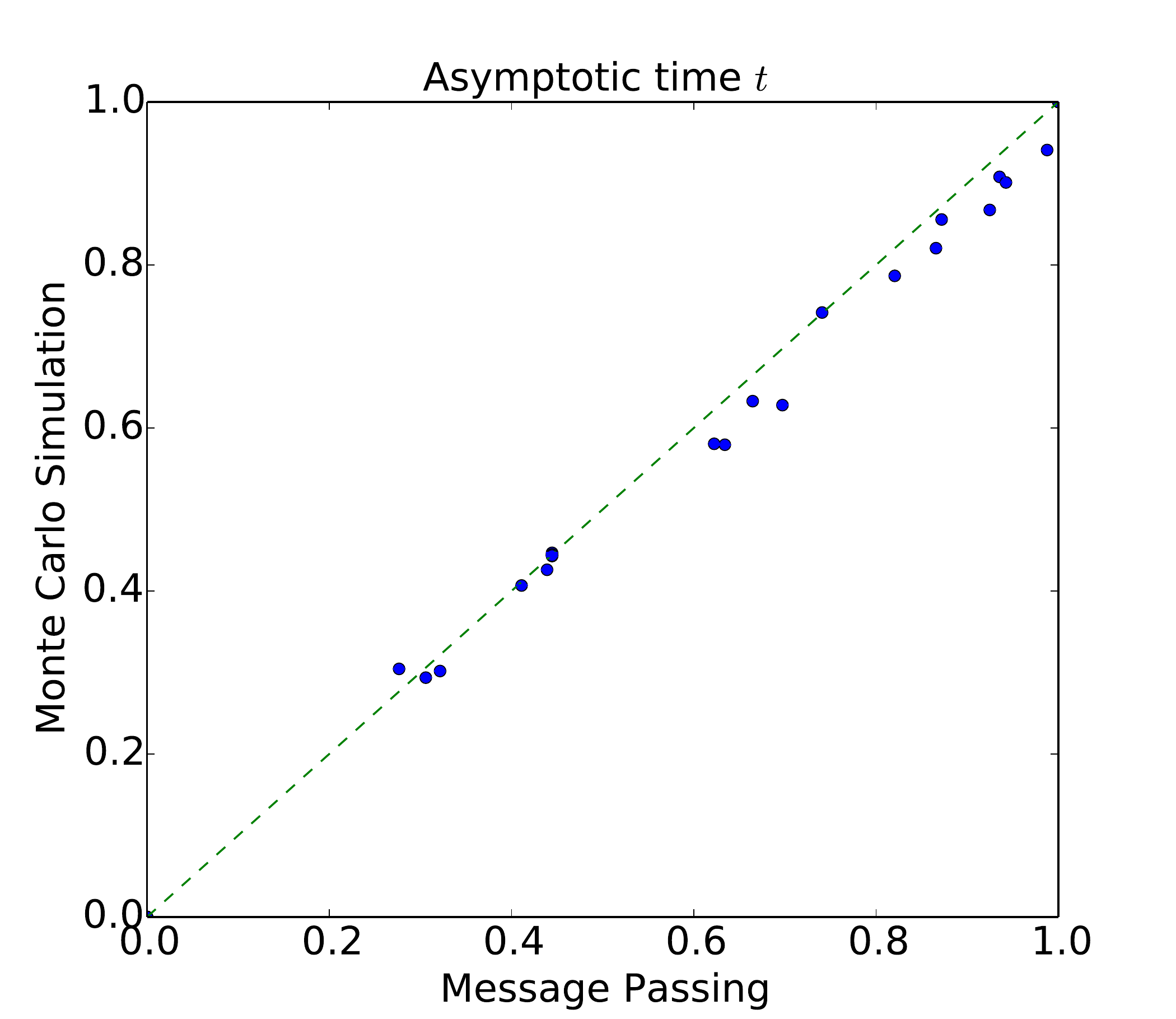} }}
\caption{Comparison (right) with a scatter plot of individuals eventual infection probability in the Zachary club (left), where threshold $T=2$. Horizontal axis is the eventual infection probability calculated by the DMP, whereas vertical axis is the result from the Monte Carlo simulation. Each point refers to the eventual infection probability of one of the individuals in the club. Here, four initially infected individuals are $\{0, 1, 32, 33 \}$. Simulation is averaged over $10^5$ runs. Transmission rate $\beta =$ 0.6, and recovery rate $\gamma =$ 0.3. Vertices on the left are colored according to their eventual infection probability from the DMP.}
\label{DMPvsABS}
\end{figure*}

\begin{figure*}
\centering
\mbox{\subfigure{\includegraphics[width=3.5in]{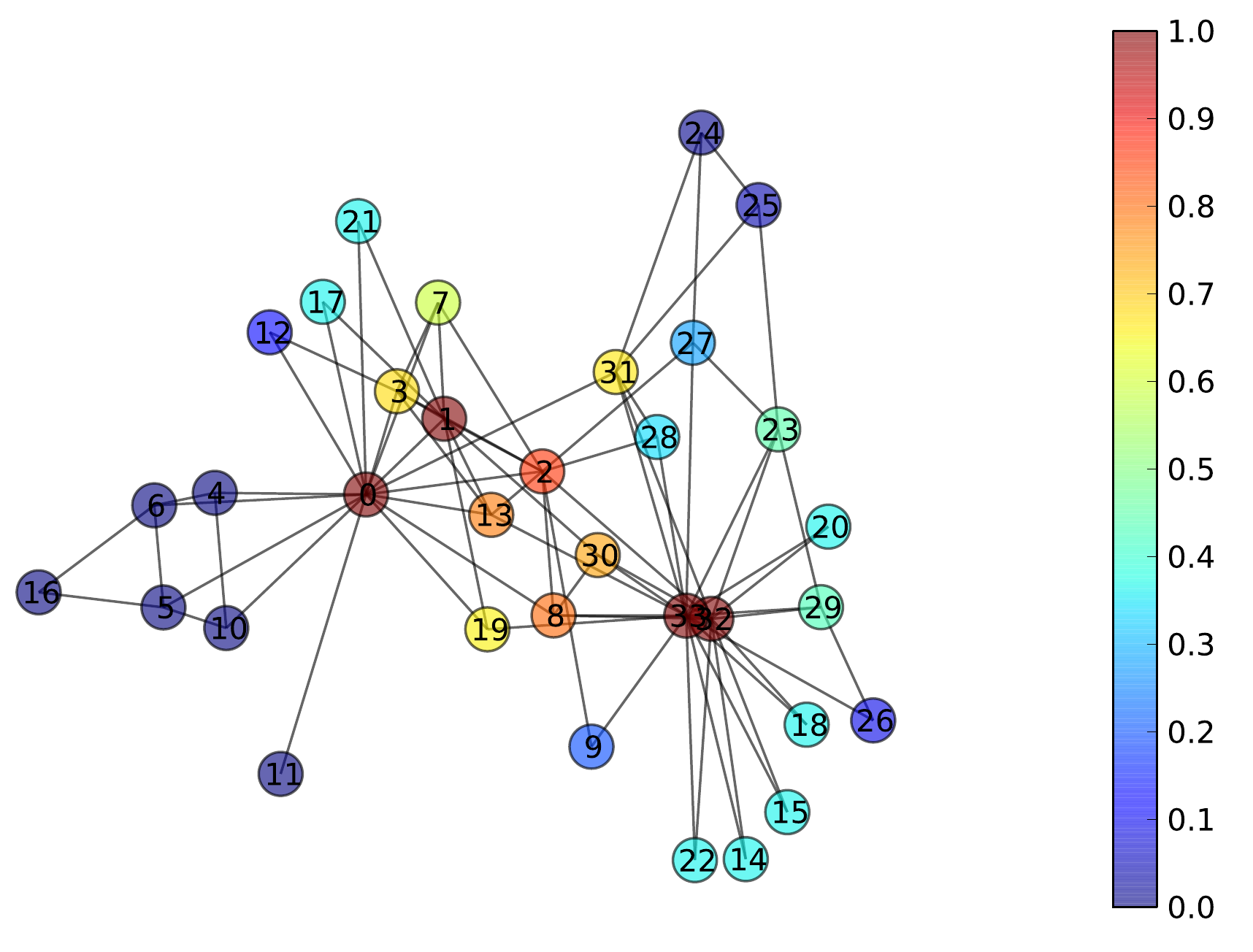}}\quad
\subfigure{\includegraphics[width=3in]{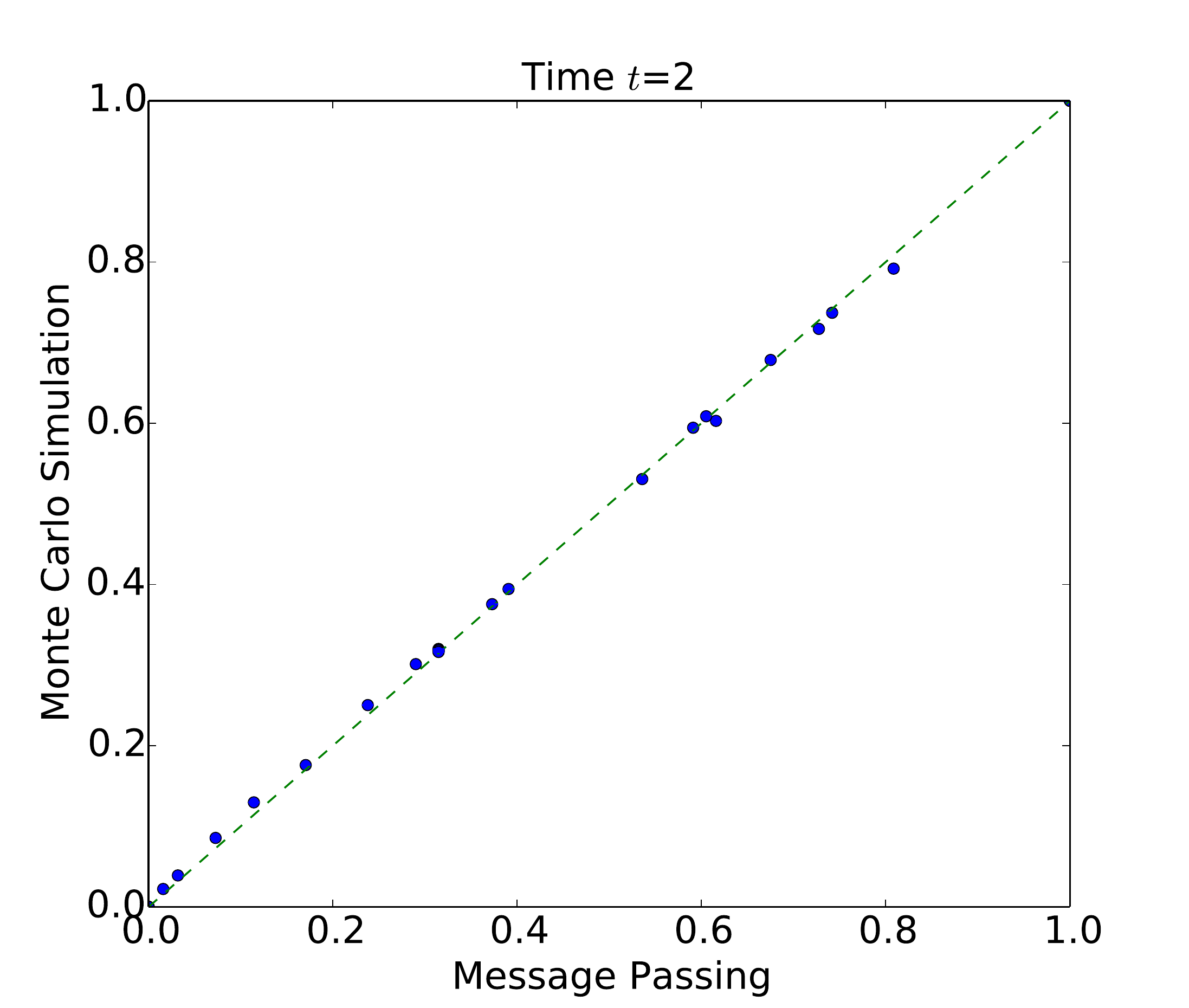} }}
\caption{Same parameters and initial conditions as Fig.~\ref{DMPvsABS}, except that we are comparing the infection probability at time $t=2$.}
\label{DMPvsABS_finiteT}
\end{figure*}

\begin{figure*}
\centering
\mbox{\subfigure{\includegraphics[width=3.5in]{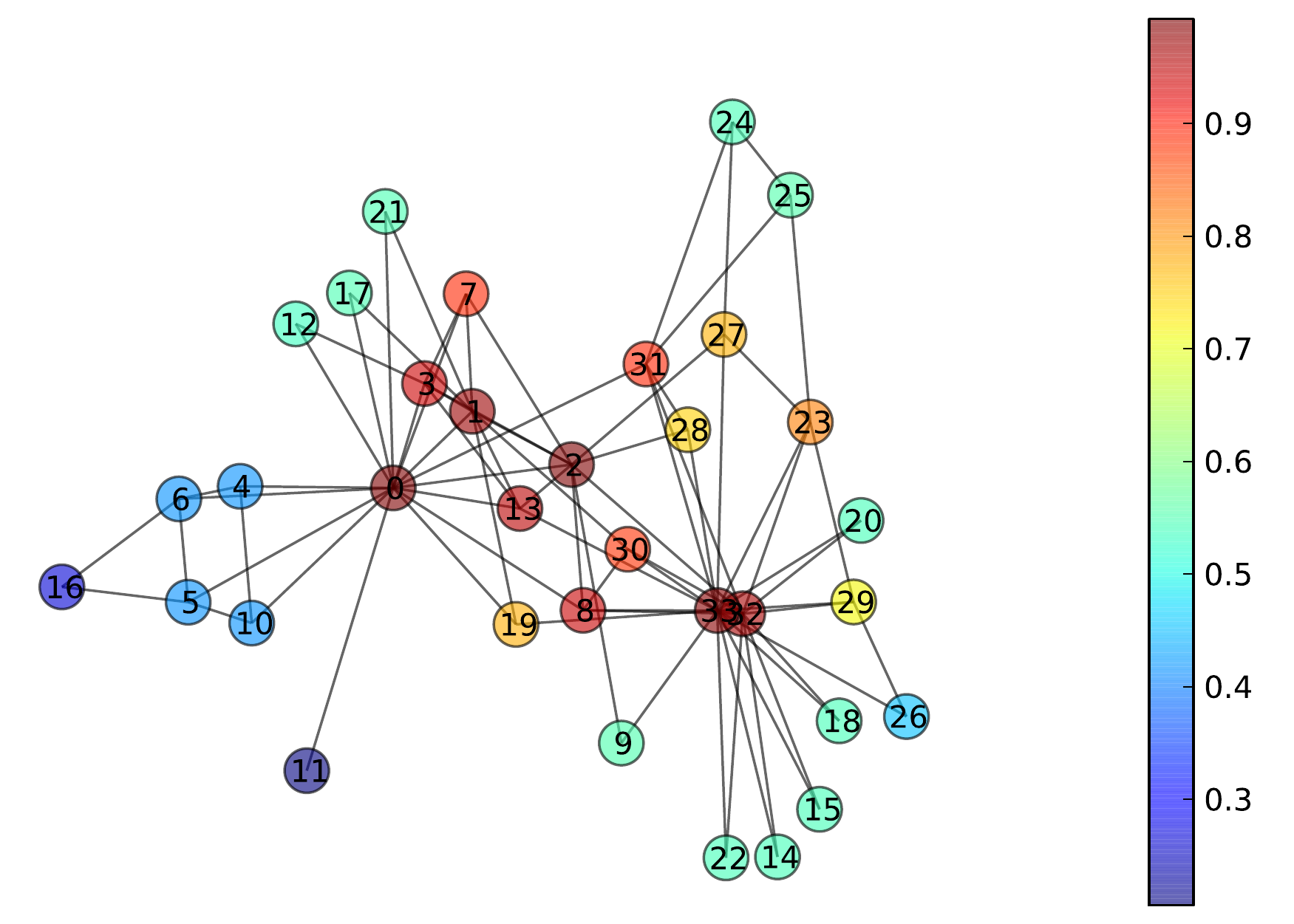}}\quad
\subfigure{\includegraphics[width=3in]{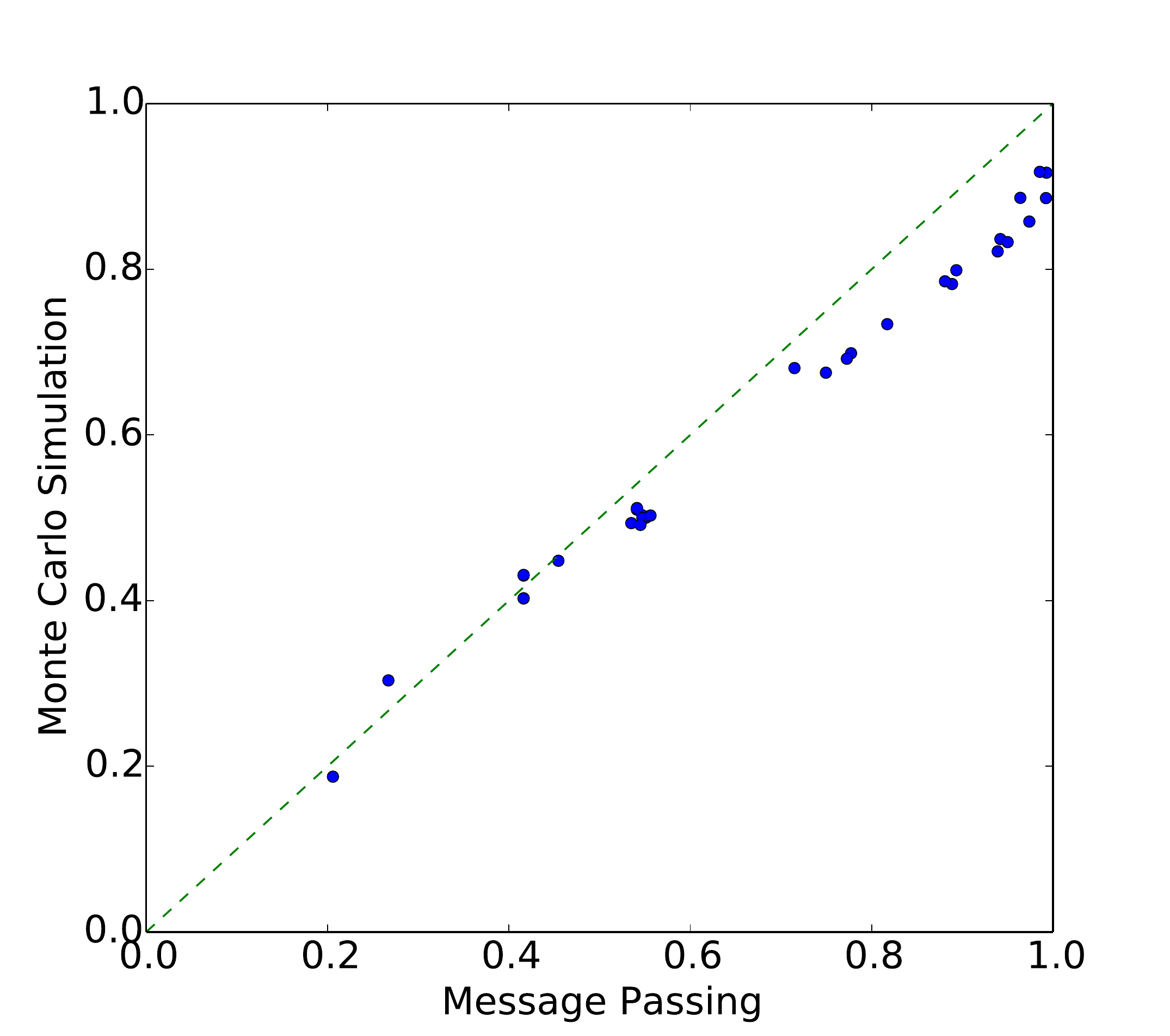} }}
\caption{Same as Fig.~\ref{DMPvsABS}, where we compare individuals probability of eventually getting infected. Here the initial condition is such that each is infected with probability 0.2.  }
\label{DMPvsABS_phi}
\end{figure*}

\begin{figure*}
\centering
\mbox{\subfigure{\includegraphics[width=3.5in]{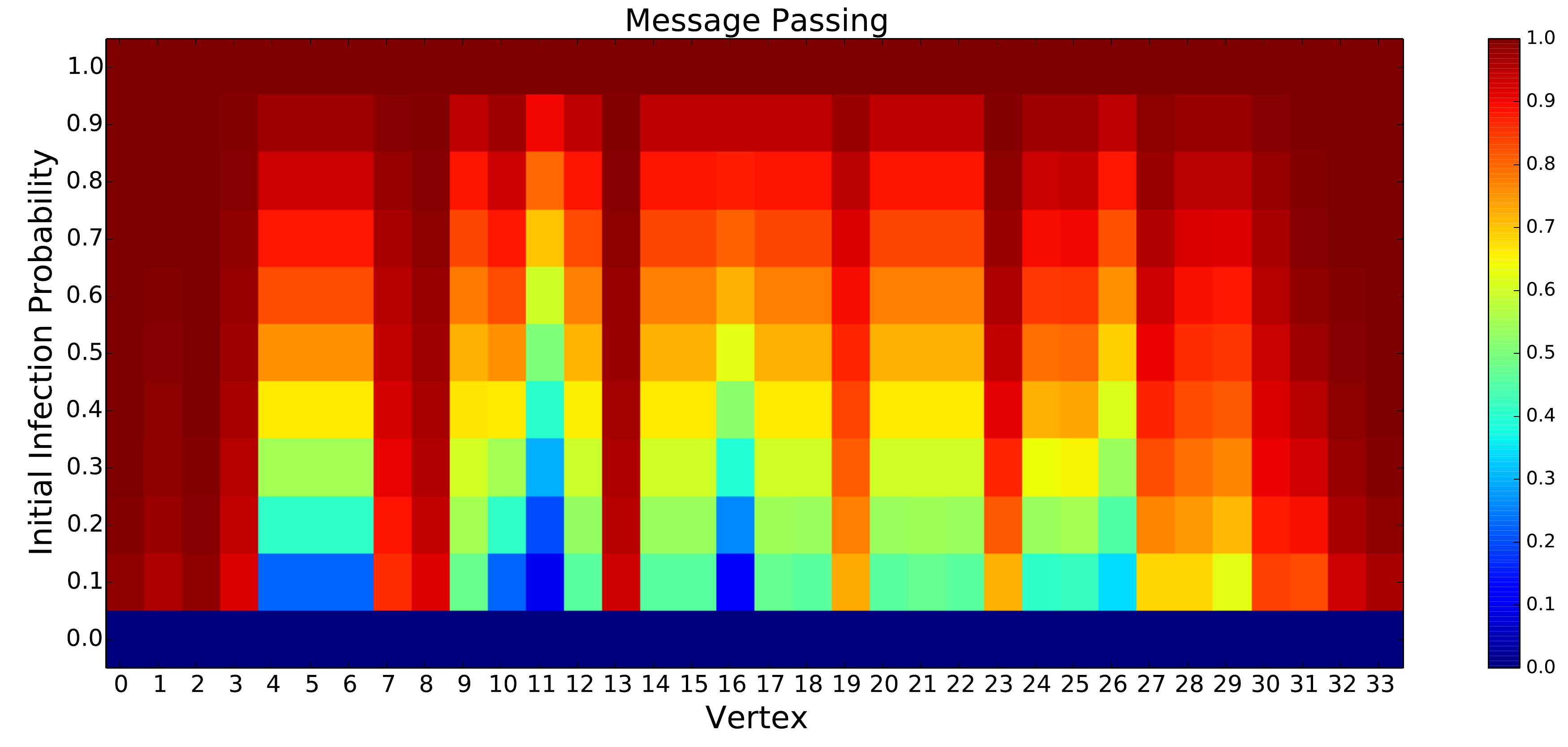}}\quad
\subfigure{\includegraphics[width=3.5in]{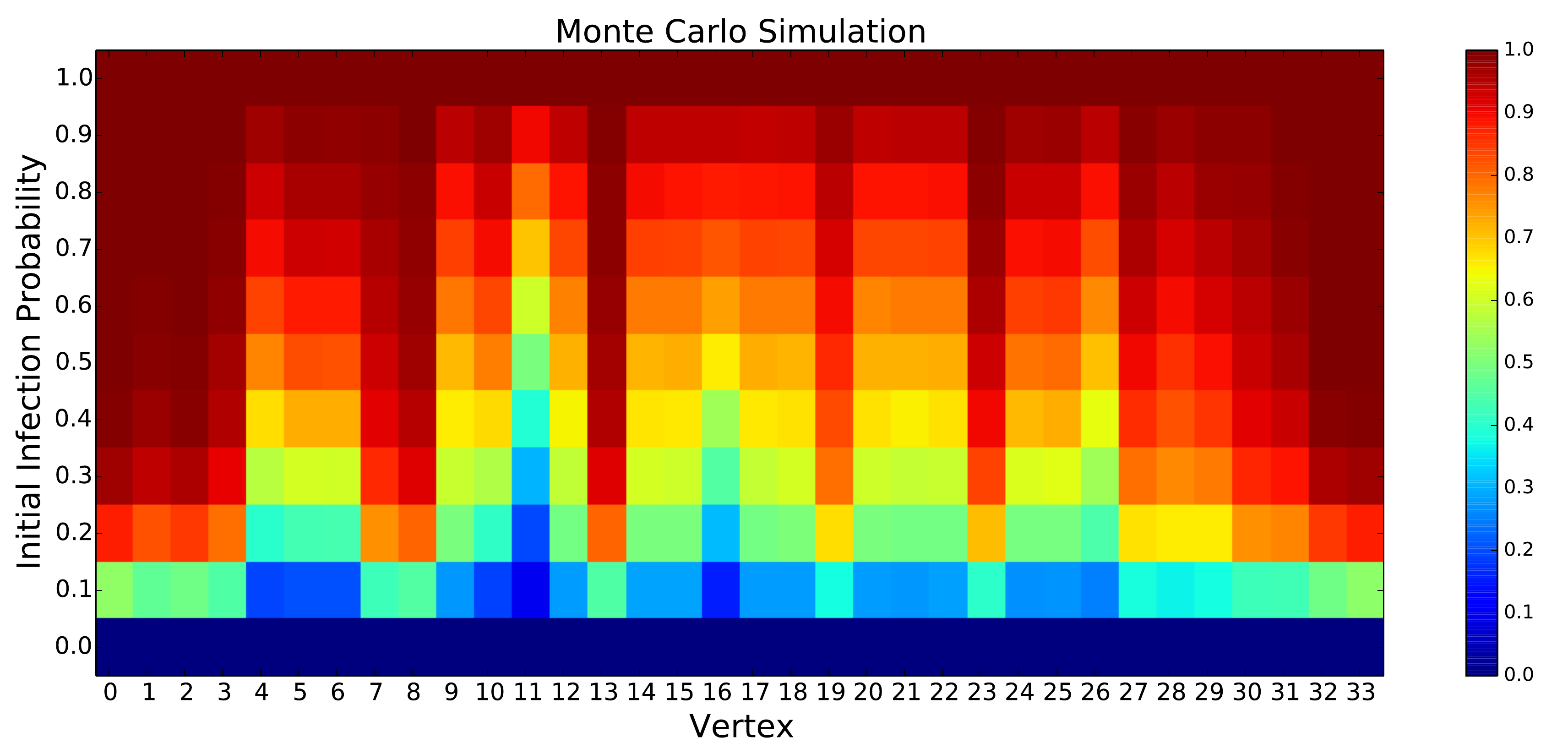} }}
\caption{We show the eventual infection probability of each individual (horizontal axis) in the Zachary karate club network at increasing uniform probability (vertical axis) of getting infected initially. Here, threshold $T =2$, transmission rate $\beta = 0.6$, and recovery rate $\gamma = 0.3$. On the left is the result calculated through the DMP. Whereas, on the right, we show the result from the Monte Carlo simulations, where the probabilities are averaged over $10^5$ runs for each initial infection probability.}
\label{fig_phi_DMP_ABS}
\end{figure*}

The message passing formulation in Section \ref{message_passing} is exact only on trees, since we assumed that the probabilities $P_a^i(t)$ are independent. However, typical networks contain many loops. Thus, the independence assumption of the message passing approach is an approximation in real networks. Our goal in this section is to see how accurate DMP is in real networks by comparing it with Monte Carlo simulations of the actual stochastic process.  

To compare the results between DMP and Monte Carlo simulations, we show the infection probability of each individual calculated through both methods in a scatter plot. In Fig.~\ref{DMPvsABS}, we compare the eventual infection (adoption) probability of each individual in Zachary's karate club network.  Each point in the scatter plot refers to the eventual infection probability of an individual in the club. If the DMP were exact, all points in the figure would lie exactly on the dotted diagonal line. 

Here, each individual's threshold $T$ is set to 2. Four vertices labeled $\{ 0,1,32,33 \}$ in Fig.~\ref{DMPvsABS} (left) are the initially infected individuals. We assume $f(\tau)=\beta e^{-(\gamma+\beta) \tau}$ with a transmission rate $\beta=0.6$ and a recovery rate $\gamma=0.3$. We simulate the actual stochastic process using a continuous-time Monte Carlo method algorithm. Events are maintained in a priority queue using a heap data structure to sort the events in the model: specifically, sort the edges $(i,j)$ according to the time at which $j$ will inform $i$. The probabilities are then averaged over $10^5$ independent runs. 

In Fig.~\ref{DMPvsABS_finiteT}, using the same parameters and initial conditions as Fig.~\ref{DMPvsABS}, we compare the infection probability of each individual at a particular finite time $t=2$. We chose this time because this is when the average number of infected individuals is at its maximum. 

In Fig.~\ref{DMPvsABS_phi}, we again use the same parameters as Fig.~\ref{DMPvsABS}, but with different initial conditions. Each individual is initially infected with probability $0.2$. There are now two sources of randomness in the model: the dynamics and the set of initial adopters.  This again forces us to do many independent runs of the Monte Carlo simulation to estimate the infection probabilities. By setting $P_S^i(0)$ = 0.8 in Equation $\eqref{PSi}$ however, we can calculate the infection probability with the same computational cost as before where the initial infectors were fixed. Accordingly in Fig.~\ref{fig_phi_DMP_ABS}, we show the density plot of the probability that each individual (horizontal axis) is eventually infected, when each of them is initially infected with increasing probability (vertical axis).  

Checking the scatter plot of the results computed from DMP and Monte Carlo simulation in Figures \ref{DMPvsABS} - \ref{DMPvsABS_phi}, we first see that the results computed from DMP do not match perfectly with those from the simulation. As pointed out in \cite{KarrNew1}, where $T=1$ the probability estimated by DMP is always an upper bound on the true probability, since the events that two or more neighbors become infected are positively correlated. 

However, for $T > 1$ the situation is more complicated, and DMP does not necessarily give an upper bound on the infection probability. Indeed, in Figs.~\ref{DMPvsABS}--\ref{DMPvsABS_phi}, we see several cases when DMP underestimates the infection probability rather than overestimating it. This includes the vertices labeled \{26\} in Fig.~\ref{DMPvsABS}, $\{12, 26, 27, 28\}$ in Fig.~\ref{DMPvsABS_finiteT}, and $\{5, 6, 16\}$ in Fig.~\ref{DMPvsABS_phi}. 

To see why this happens, suppose $i$ has two neighbors, $j$ and $k$.  Let $P[i]$ denote the probability that $i$ becomes infected, and let $P[j]$ and $P[k]$ denote the probabilities that $j$ and $k$ inform $i$ respectively.  
If $T=1$, then
\[
P[i] 
= P[j \vee k] 
= P[j] + P[k] - P[j \wedge k].
\]
Let's assume that DMP computes the right marginals, so that $\PDMP[j]=P[j]$ and $\PDMP[k] = P[k]$.  However, DMP ignores correlations, and assumes that these events are independent.  Thus
\[
\PDMP[i]
= P[j] + P[k] - P[j] P[k] .
\]
However, $j$ and $k$ are positively correlated if they have a common neighbor that may have infected them both, or if they are neighbors of each other.  That is, 
\[
P[j \wedge k] > P[j] P[k].
\]
Then $P[i] < \PDMP[i]$, and DMP overestimates $P[i]$.  
On the other hand, if $T=2$, then 
\[
P[i]
= P[j \wedge k]
> P[j] P[k]
= \PDMP[i], 
\]
and DMP underestimates $P[i]$.  

Similarly, suppose $i$ has three neighbors, $j$, $k$, and $\ell$.  Again taking $T=2$, we have
$$
P[i]=P[j \wedge k] +P[j \wedge \ell] +P[k \wedge \ell]  -2P[j \wedge k \wedge \ell],
$$
 whereas, DMP gives
$$
\PDMP[i]=P[j] P[k] +P[j] P[\ell] +P[k] P[\ell] -2P[j]P[k]P[\ell].
$$
In this case, DMP can either underestimate or overestimate $P[i]$, depending on the strength of the correlations between its neighbors.  For example, if $\ell$ is independent of $j$ and $k$, then
\begin{align*}
P[i]&=P[j \wedge k] +P[j] P[\ell] +P[k] P[\ell]  -2P[j \wedge k] P[\ell] \\ 
& =P[j \wedge k] (1-2P[\ell]) + (P[j]+P[k]) P[\ell] .
\end{align*}
If $j$ and $k$ are positively correlated so that $P[j \wedge k] > P[j] P[k]$, then DMP underestimates $P[i]$ if $P[\ell] <1/2$ and overestimates it if $P[\ell] >1/2$.


\section{Exact Solution in Networks with arbitrary degree distributions}
\label{configuration_network}

In this section, we consider the message passing approach in the ensemble of random networks in the thermodynamic limit. Our goal is to show that DMP can be applied to large random networks just as well as to a particular finite network. 

In random networks, we are interested in the expected behavior of the dynamics rather than the dynamics in a single realization of the network. So, instead of computing messages for individual vertices, we assume that these messages are drawn from some probability distribution, and update this distribution based on their average behavior. We can then compute the distribution of marginals as well.

We consider random networks with a given degree distribution, specifically an ensemble of networks called the \emph{configuration} model \cite{NewWatt1}. Each of $n$ vertices is first assigned an integer degree from a specified degree distribution, say $p_k$. We think of a vertex with degree $k$ as having $k$ ``spokes" or half-edges coming out of it.  We then choose a uniformly random matching of these $2m$ spokes with each other, where $m$ is the number of edges in the network. The key fact is then that, in the thermodynamic limit, i.e. $n \to \infty$, following an edge from any given vertex connects with a vertex of degree $k$ with probability proportional to $kp_k$. Strictly speaking, this model generates random multigraphs. But, the average size of such graphs is a constant as $n \rightarrow \infty$, as a result of which the density of self-loops and multiple edges vanishes when $n$ is large. 

Now, consider the message $U_{i\leftarrow j}(t)$ from Equation \eqref{mainUij}. Recall that this is the probability that $j$ has not informed $i$ by time $t$. In the configuration model however, different individuals $j$ are connected to $i$ in different realizations of the network. But, edges are now statistically identical in the sense that each edge identically connects to a vertex based on its degree. So, we consider a single average message $U(t)$.   

This average message $U(t)$ then has the following interpretation. It is the average probability that by following a random edge, the neighbor we reach has not informed the vertex we came from by time $t$. This in turn will tell us the probability $P_{a}(t)$ that a randomly chosen vertex has awareness $a$ at time $t$. However, this probability depends on the degree of the vertex: specifically, if it has degree $k$, then 
\be
P_a(k, t) = P_S(0) {k\choose a} U(t)^{k-a} (1-U(t))^a.
\ee
Averaging over $p_k$, we get 
\be
P_a(t) = P_S(0) \sum_{k}^\infty p_k  {k\choose a} U(t)^{k-a} (1-U(t))^a.
\ee
It is useful to write this in terms of the generating function $G(x)$ of the degree distribution and its derivatives:
\begin{align*}
&G(x)= \sum_k p_k x^k, \numberthis  \\
&G^{(a)}(x)=\frac{d^a G(x)}{dx^a}. \numberthis
\end{align*}
Then $P_{a}(t)$ can be written as
\be\label{gen_pa}
P_{a}(t) =P_S(0)\frac{(1-U(t))^a}{a !} G^{(a)}(U(t)). 
\ee
Thus the probability $P_{S}(t)$ that a randomly chosen vertex is susceptible at time $t$ is 
\be
P_{S}(t)= \sum_{a=0}^{T-1} P_{a}(t).
\ee
Equivalently, 
\be\label{gen_susp}
P_{S}(t) =P_S(0)\sum_{a =0}^{T-1} \frac{(1-U(t))^a}{a !} G^{(a)}(U(t)). 
\ee
So, we see that given $U(t)$, computing $P_a(t)$ and $P_S(t)$ in the configuration model reduces to knowing $G^{(a)}$ to some order. 

To capture the information flow that $U(t)$ represents in the configuration model, we define the cavity probability $Q(t)$ by simplifying Equation \eqref{cavity_PS}. This is the probability that a randomly chosen edge leads to a vertex that has $not$ been infected by time $t$, if the vertex we came from is assumed to be absent from the network. Equivalently, $Q(t)$ is the probability that if we follow a random edge from a vertex $i$, the vertex $j$ it leads to has been informed by at most $T-1$ of its neighbors other than $i$. This probability also depends on $j$'s degree. Namely, if it has degree $k+1$, then  
\be\label{mainZ}
Q (k,t) = \sum_{a =0}^{T-1}{k\choose a} U(t)^{k-a} (1-U(t))^a, 
\ee
where $k$ is the number of neighbors that $j$ has other than $i$. As discussed above, a random edge leads to a vertex with degree $k$ with probability proportional to $k p_k$. Therefore, the probability that $j$ has $k$ neighbors other than $i$ is
\be 
q_{k} = \frac{(k+1)p_{k+1}}{\sum_k k p_k}=\frac{(k+1)p_{k+1}}{G^{(1)}(1)}.
\ee
Averaging $Q (k,t)$ over $q_k$, we obtain  
\begin{align*}
Q (t) &= \sum_k q_k \sum_{a =0}^{T-1}{k\choose a} U(t)^{k-a} (1-U(t))^a. \numberthis
 \end{align*}
Similar to Equation \eqref{gen_susp}, we can write $Q (t)$ in terms of the generating function as 
\begin{align*}
Q (t) &= \frac{1}{G^{(1)}(1)}\sum_{a =0}^{T-1} \frac{(1-U(t))^a}{a !} G^{(a+1)}(U(t)). \numberthis
 \end{align*}

We now calculate $U(t)$ by simplifying (i.e. averaging) Equation \eqref{mainUij} for the configuration model. But, note the right-hand side of \eqref{mainUij} consists of products of $U(t)$, and the average of products is not always the product of averages. In the limit $n\rightarrow \infty$ however, the network is locally treelike in the sense that the typical size of the shortest loops diverges as $O(\log n)$. As a result, $U(t)$ is asymptotically independent, and the average of products is equal to the product of averages. So, the self-consistent relation for $U(t)$ becomes  
\begin{align*}
 U(t) = 1- \int_0^t d\tau f(\tau) +P_S(0) \int_0^t dt' f(t-t') Q(t') \numberthis  \label{mainU}.
\end{align*}
To numerically integrate this equation in time, we differentiate it with respect to $t$,
\begin{align*}
 \frac{dU(t)}{dt} = - f(t) &+ P_S(0)f(0) Q(t) \\ &+ P_S(0) \int_0^t dt'Q(t')  \frac{d f(t-t')}{dt}.   \numberthis \label{mainUdiff2}
\end{align*}
It is also possible to get this from Equation \eqref{mainUijdiff2}. We can further simplify this to an ordinary differential equation in some cases. For example, if $f(\tau)= \beta e^{-(\beta+\gamma)\tau}$, we can write it as 
\begin{align*}
 \frac{dU(t)}{dt} = -\beta U(t)+\gamma (1- U(t))+ \beta P_S(0) Q(t)     \numberthis \label{mainUdiff3}.
\end{align*}

\begin{figure*}
\centering
\mbox{\subfigure{\includegraphics[width=3.5in]{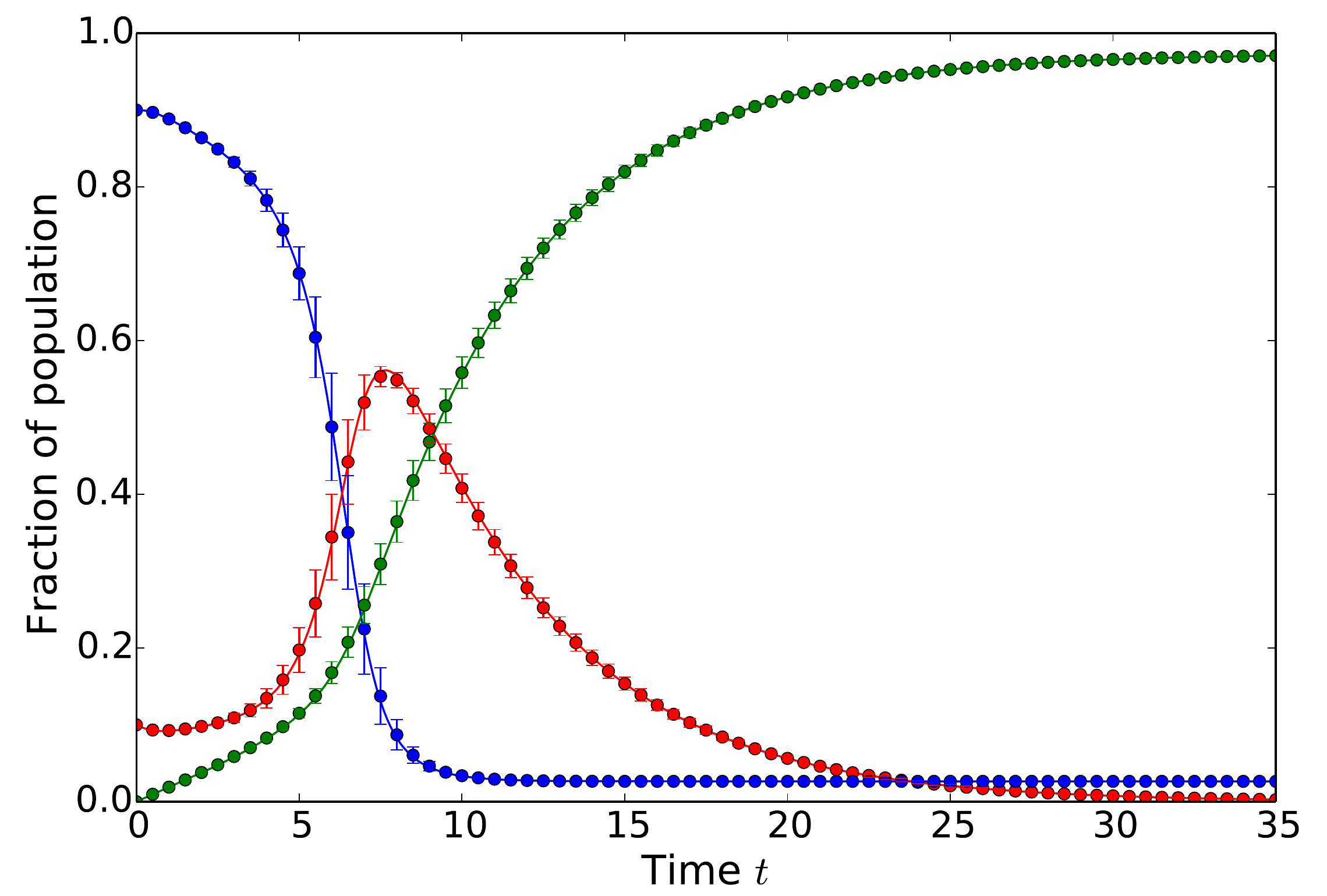}}\quad
\subfigure{\includegraphics[width=3.5in]{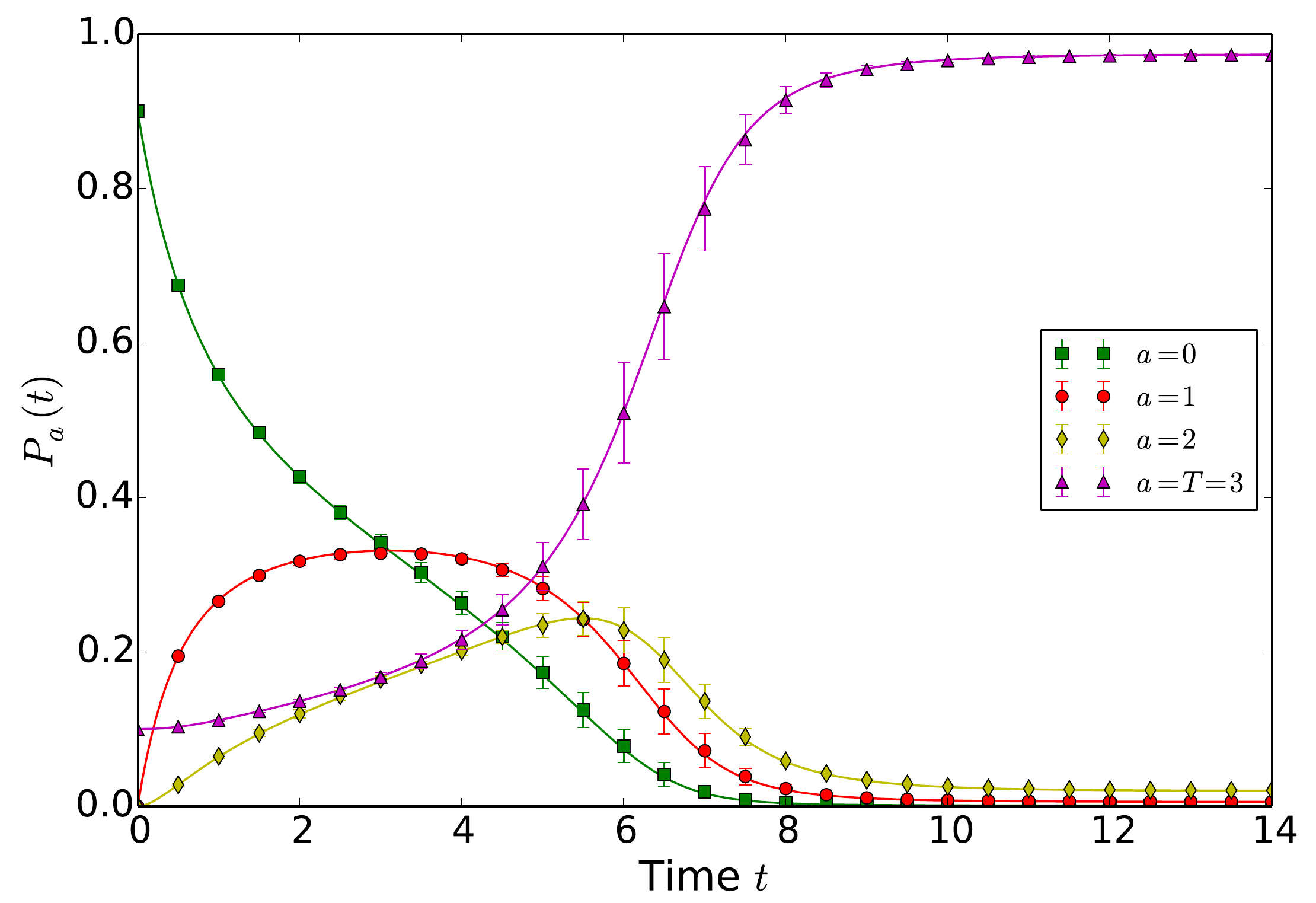} }}
\caption{On the left is the dynamics in the Erd\H{o}s-R\'enyi graphs $G(n,p=c/n)$ where individuals have threshold $T=3$, average degree $c=9$, initial fraction of adopters/infecteds $P_T(0)$ = 0.1. The fractions of infected, recovered and susceptible vertices are red, green, blue respectively. Continuous lines are analytic results calculated using our DMP approach, by numerically integrating Equation \eqref{mainUdiff3}, whereas dots are from the Monte Carlo based simulations with $10^4$ vertices averaged over 100 runs. Transmission rate $\beta =$ 0.8, and recovery rate $\gamma =$ 0.2. On the right is the time evolution of $P_a(t)$, where continuous lines are calculated using Equation \eqref{gen_pa}. Root Mean Square deviations in the simulation are provided when they are larger than the markers.}
\label{config1}
\end{figure*}

\begin{figure}
\centering
\includegraphics[width=3.5in]{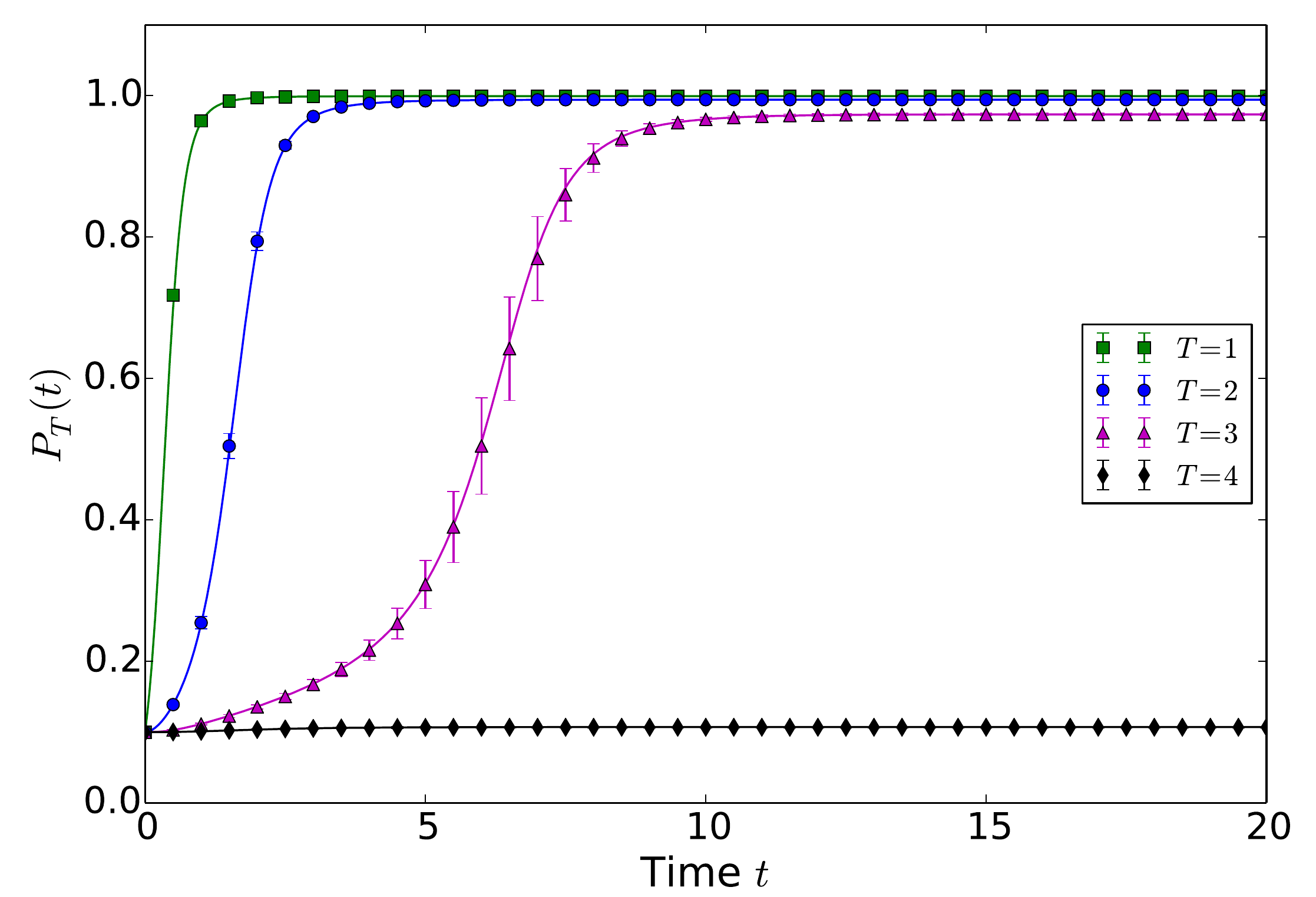}
\caption{Same parameters and initial conditions as Fig.~\ref{config1}, except we are computing the fraction $P_T(t)$ of adopters, i.e. either infected or recovered vertices,  as a function of time when the threshold $T$ is 1 (green square), 2 (blue circle), 3 (magneta triangle), and 4 (black diamond).}
\label{PS_T_config}
\end{figure}

So, given the initial conditions $U(0)=1, P_S(0)$, and $G^{(a)}(x)$, we can calculate $P_S(t)$ using Equation \eqref{gen_susp}. Similarly, the fraction of infected and recovered vertices at time $t$ can be calculated. Note that, in general, we can let $f(\tau)$ depend on the degree of the vertex by following a degree dependent transmission method formulated by Newman \cite{Newman1}. Similarly, we can allow for the case where the probability $P_T(0)=1-P_S(0)$ of getting initially infected depends on the degree of the vertex. 

In Fig.~\ref{config1} (left), we show the time evolution of the fraction of susceptible (blue), infected (red), and recovered (green) vertices in the configuration model, where the degrees are drawn from the Poisson distribution with mean $c$, or equivalently the Erd\H{o}s-R\'enyi graphs $G(n,p=c/n)$. For Poisson distribution, $G^{(a)}(x)$  are given by $c^a e^{-c(1-x)}$. We take $c=9$, $T=3$, $f(\tau)=\beta e^{-(\beta+\gamma)\tau}$ , where  $\beta = 0.8$ and $\gamma =0 .2$, and the initial fraction of adopters/infecteds is $P_T(0)= 0.1$. 

Continuous lines in Fig.~\ref{config1} (left) are obtained by numerically integrating Equation \eqref{mainUdiff3},whereas dots are from Monte Carlo simulations with $10^4$ vertices averaged over 100 runs. Similarly, Fig.~\ref{config1} (right) gives the fraction $P_a(t)$ of vertices with awareness $a$, where the continuous lines are obtained by using Equation \eqref{gen_pa}. 

In Fig.~ \ref{PS_T_config}, we show the fraction $P_T(t)$ of adopters  as a function of time for the same parameter values as Fig.~\ref{config1}, except where $T$ is 1 (green square), 2 (blue circle), 3 (magneta triangle), and 4 (black diamond). Root Mean Square deviations in the simulation are provided when they are larger than the the markers. 

Using the same framework, we can calculate the asymptotic probability $u=U(\infty)$ that the infection has not been transmitted along a random edge. This in turn will tell us the asymptotic probability that a randomly chosen vertex ever becomes infected. 

We can think of the long time behavior as $k$-core percolation.  Either the edge is closed in the sense that its other endpoint fails to inform the vertex we came from, which happens with the probability $1-p=1-\int_0^\infty f(\tau)d\tau$. In this case, it does not matter if the neighbor gets infected by its other neighbors, since it fails to inform the vertex we came from. Or, it can be the case that the edge is open (with probability $p$), but the vertex we reach is itself not infected eventually by its other neighbors. This happens when the neighbor we reach by randomly following the edge is informed by at most $T-1$ other neighbors, provided it was not initially infected. Summing up both cases, we arrive at the following self-consistent relation for $u$: 
\begin{align*}
u&=  1-p +p P_S(0)\sum_{k}^\infty q_k \sum_{a =0}^{T-1}{k\choose a}  u^{k-a} (1-u)^a \\
&=1-p +\frac{p  P_S(0)}{G^{(1)}(1)} \sum_{a =0}^{T-1} \frac{(1-u)^a}{a !} G^{(a+1)} (u)
 .\numberthis \label{longtimeU}
\end{align*}
Note that we could have written this equally by taking the limit $t \to \infty$ in Equation \eqref{mainU}. Similarly, the probability $P_S$ that a randomly chosen vertex never gets infected, i.e.  the fraction of susceptible vertices is
\be\label{PS_longtime}
P_S= P_S(0)\sum_{a =0}^{T-1} \frac{ (1-u)^a}{a !} G^{(a)} (u) .
\ee
For Erd\H{o}s-R\'enyi networks $G(n,p=c/n)$, or equivalently the Poisson distribution with average degree $c$, we have the following self-consistent relation for $u$:
\be
u=  1-p +p P_S(0)e^{-c(1-u)} \sum_{a =0}^{T-1} \frac{c^a(1-u)^a} {a !}.
\ee   
We can also obtain this expression by following \cite{BaxDorog}. Similarly, $P_S$ in Erd\H{o}s-R\'enyi networks is
\be
P_S= P_S(0)e^{-c(1-u)} \sum_{a =0}^{T-1} \frac{c^a(1-u)^a} {a !}.
\ee

Equations \eqref{longtimeU} and \eqref{PS_longtime} have a nice interpretation in terms of well-studied problems in random graphs, including percolation and the emergence of the $k$-core. We say that Equation \eqref{longtimeU} is the generating function in $P_S(0)$ of the size of the connected component of susceptible vertices by following a random edge in the long time limit. Similarly, Equation \eqref{PS_longtime} is the generating function of the size of the connected susceptible component of a randomly chosen vertex. 

\section{Concluding Remarks and Generalizations}

In this paper, we have considered the dynamic message-passing (DMP) technique to study a simple threshold model of behavior in networks. In doing so, we are able capture how each individual's probability of becoming an adopter evolves in time in an arbitrary network with far less computational cost than Monte Carlo simulations.  Although DMP is exact only on trees, we observe that it compares well with simulations even in a real social network where there are many loops. Interestingly, unlike in the SIR model, or equivalently the case $T =1$, there are cases where DMP can either underestimate or overestimate the probability of infection. 

In addition, we have used the DMP equations to give analytical results in the thermodynamic limit of large random networks. We have provided an exact analytic result for calculating the time dependence of the probabilities, thereby learning something about the dynamics of bootstrap percolation.

The message-passing dynamics we have considered here can be generalized in many ways, including letting the transmission probability and the threshold vary arbitrarily across edges and vertices. Because the transmission rate $r(\tau)$ may depend on the elapsed time $\tau$ since an individual became an adopter, our study can be implemented in networks where some non-Markovian assumptions are warranted, as we pointed out in Section \ref{message_passing}. 

We can include so-called ``rumor spreading'' models where, rather than setting $r(\tau)=0$ until an individual's awareness reaches a threshold as we have done here, an individual starts telling its neighbors about the rumor even if it has only heard about it once. Such models were recently applied to the diffusion of microfinance \cite{MJackson}. We can also let the rate at which an individual receives new information depend on its own awareness. An interesting case is to consider a unimodal function. 

We can also consider a model where $j$ can transmit repeatedly to $i$, raising $i$'s awareness each time.  We simply replace each directed edge $(j,i)$ with $T$ multi-edges. So, each message $U_{i\leftarrow j} (t)$ would now be mapped to $T$ identical copies of itself. The update equations and expressions are the same as above, but now we sum over all these multi-edges accordingly.

In Section \ref{configuration_network}, we focused on random networks in the configuration model.  However, the DMP equations can be easily generalized to many other families of random graphs, including interdependent networks \cite{Havlin_Stanley_Interdependent}, scale-free networks \cite{Barabasi99}, small-world networks \cite{WS98, MN00a}, and bipartite networks \cite{CSMF2012} to name a few. In some cases this is a matter of plugging in a different degree distribution, and allowing for a finite number of types of vertices.  However, for preferential attachment networks the topology is correlated with the vertices' ages, so we would have to let the messages $U(t)$ depend on the age of the vertices sending them.

We can also extend this study to a network that has community structures such as the stochastic block model. We can then study how trends move through communities, and how the distribution of initial adopters (for instance, whether they are concentrated in one community, or are spread across many communities) affects the eventual fraction of the network that adopts the trend. Community structures can be driven by socio-economic, ethnic, religious and linguistic separations. So, it would be useful to gain some perspective on how the structures of communities contribute to the norms and social preferences that prevail in real populations, and in turn how differences in these norms drive the division of social networks into communities. 

\section{Acknowledgments}
M.S. is supported by National Institute of Health under grant \#T32EB009414. M.S. and C.M. are supported by the AFOSR and DARPA under grant \#FA9550-12-1-0432. The authors would like to thank Mark Newman, Brian Karrer, and Pan Zhang for useful discussions.


\end{document}